# Rapid three-dimensional multiparametric MRI with quantitative transient-state imaging

**Pedro A. Gómez [1], Matteo Cencini [3,4], Mohammad Golbabaee [5], Rolf F. Schulte [6], Carolin Pirkl [1,6], Izabela Horvath [1,6], Giada Fallo [2,3], Luca Peretti [2,3], Michela Tosetti [3,4], Bjoern H. Menze [1], Guido Buonincontri [3,4]**

[1] Computer Science, Technical University of Munich, Munich, Germany

[2] University of Pisa, Pisa, Italy

[3] Imago7 Foundation, Pisa, Italy

[4] IRCCS Stella Maris, Pisa, Italy

[5] University of Bath, Bath, UK

[6] GE Healthcare, Munich, Germany

Corresponding author:

Dr. Pedro A. Gómez

Munich School of Bioengineering

Technical University of Munich

Boltzmannstr. 11, D-85748 Garching, Germany

E. pedro.gomez@tum.de

# Abstract

Novel methods for quantitative, transient-state multiparametric imaging are increasingly being demonstrated for assessment of disease and treatment efficacy. Here, we build on these by assessing the most common Non-Cartesian readout trajectories (2D/3D radials and spirals), demonstrating efficient anti-aliasing with a *k*-space view-sharing technique, and proposing novel methods for parameter inference with neural networks that incorporate the estimation of proton density.

Our results show good agreement with gold standard and phantom references for all readout trajectories at 1.5T and 3T. Parameters inferred with the neural network were within 6.58% difference from the parameters inferred with a high-resolution dictionary. Concordance correlation coefficients were above 0.92 and the normalized root mean squared error ranged between 4.2% - 12.7% with respect to gold-standard phantom references for T1 and T2. *In vivo* acquisitions demonstrate sub-millimetric isotropic resolution in under five minutes with reconstruction and inference times < 7 minutes.

Our 3D quantitative transient-state imaging approach could enable high-resolution multiparametric tissue quantification within clinically acceptable acquisition and reconstruction times.

# Introduction

Quantitative, multiparametric imaging offers great opportunities for assessment of disease and treatment efficacy. Compared to contrast-weighted images, quantitative measurements are less influenced by system and interpretation biases, increasing data reproducibility and accuracy[1]. However, obtaining quantitative information usually requires the acquisition of multiple views along each parameter-encoding dimension, which often leads to impractically long scan times. To overcome this challenge, various rapid multiparametric mapping techniques have recently been demonstrated[2–5]. Compared to methods for estimating a single parameter at a time, the models at the heart of these techniques have the main advantage of accounting for the correlations between all quantitative parameters of interest as well as system imperfections. This holistic view of the MR signal enables the simultaneous regression of many individual parameters, potentially with higher accuracy. Novel, accurate techniques promise a fast estimation of relevant MRI quantities, including but not limited to, the longitudinal (T1) and transverse (T2) relaxation times, allowing also retrospective synthesis of conventional MR contrasts[3]. Amongst these emerging methods, MR Fingerprinting (MRF)[2], quantitative transient-state imaging (QTI)[4], and Magnetic Resonance Spin TomogrAphy in Time-domain (MR-STAT)[5], are currently being demonstrated in healthy subjects and/or patient groups.

These quantitative techniques, following the seminal paper by Ma et al.[2], use similar strategies to simultaneously sample the *k*-space and parameter space, relying on transient-state[6] acquisitions achieved by continuous variations of the acquisition parameters while acquiring undersampled *k*-space snapshots after each excitation. Parametric maps are subsequently computed by enforcing consistency with a physical model, i.e. the Bloch equations. While these methods have been shown to be robust to aliasing artifacts, extreme undersampling factors may still lead to quantification errors and decreased image quality. The local quantification accuracy depends both on the used flip angle schedule and the *k*-space sampling trajectory, as time-dependent point spread functions (PSF) will interfere differently in different spatiotemporal coordinates[7]. So far, the most successful implementations of transient-state imaging have relied on trajectories that oversample the *k*-space center, such as radial or spiral acquisitions, in combination with locally-smooth schedules of flip angle and repetition times[8,9]. Despite their initial success in high-resolution parameter mapping, current quantitative techniques still suffer from limitations in acquisition, reconstruction, and parameter inference efficiency.

Multiple studies have been presented to address acquisition shortcomings. For example, three-dimensional readouts have been applied – either as stack-of-spirals[7,10], 3D spiral projections[11], or music-based *k*-space trajectories[12]– reaching whole-brain high-resolution parameter mapping in clinically feasible acquisition times. Other works have focused on optimizing the flip angle schedule for efficient parameter sampling using Bayesian experimental design[4] or Crámer-Rao Lower Bound[9,13,14]. These optimization works generally converge to smooth transient-state signal evolutions, enabling the replacement of brute-force dictionary matching with voxel-wise fitting[15,16], as smooth signal evolutions avoid local minima in the Bloch-response manifold[17].

Accelerated acquisitions have been combined with advanced reconstruction algorithms and parameter inference techniques to ameliorate quantification errors associated with aliasing and noise due to increased undersampling. Simple, non-iterative methods, such as sliding window reconstruction[7,11,18] or *k*-space weighted view-sharing[19,20] have demonstrated improved quantification accuracy over traditional linear matrix inversion. Moreover, iterative algorithms further reduce quantification errors by imposing low-rank subspace constraints on either temporal signals or spatial image models[4,21–25]. Advanced reconstructions, however, generally trade-off shorter scan durations against increased computing times, hindering their clinical applicability. Finally, machine learning algorithms have demonstrated improved parameter inference; for instance, by either learning the dictionary[26,27] or completely replacing it with artificial neural networks[28–30]. The latter approach also tackles dictionary discretization and size limitations, enabling higher-dimensional applications of multiparametric mapping, such as capturing the full diffusion tensor together with relaxation parameters[29].

This work focuses on 3D acquisition, reconstruction, and parameter inference for multiparametric mapping of the brain. Our contributions are summarized as follows:

1. We provide a systematic comparison of the most common non-Cartesian readout trajectories, namely 2D/3D spiral arms and radial spokes. We compare them in terms of efficiency, coverage, PSF, and impact on quantification accuracy; showing that 3D spirals achieve the most efficient coverage for whole brain quantification.
2. To overcome computation and memory-expensive dictionary matching approaches, we employ a novel compact artificial neural network for parametric inference in a lower dimensional subspace. Along with the prediction of T1 and T2 relaxation times, we include the quantification of proton density (PD) as a linear scaling factor into the neural network setup.
3. We demonstrate a reduction in reconstruction time for 3D high-resolution mapping from 1.5 – 4 hours[11] to under 7 minutes.

## Methods

**Data acquisition and reconstruction with non-Cartesian readout trajectories**

*Sequence parameters*

Data acquisition is performed during the transient-state, with varying acquisition parameters from one repetition to the next. Repetition time (TR) and echo time (TE) were chosen to be constant and minimal throughout the experiment, resulting in TR = 10.5/12 ms (for 3T/1.5T, respectively) and TE = 0.46/2.08 ms (for spiral/radial acquisitions, respectively). The unbalanced gradient was designed to achieve a $4\pi$ dephasing over the z direction of the voxel. The choice of the variable acquisition parameters is shown in Figure 1. As described by previous work[4], the inversion pulse followed by the increasing flip angle ramp encodes simultaneously for T1 and T2[4]. The decreasing ramp will then smoothly reduce the magnetization until the last section of the flip angle train. This portion with a small and constant flip angle, in turn, allows for T1 recovery before the next inversion pulse. It also allows to repeat the acquisition obtaining full 3D encoding. We call these repetitions *segments,* as opposed to the repetitions of the excitations.

For the readout trajectory, we implemented and evaluated the most common non-Cartesian readout trajectories: 2D/3D spiral and radial readouts. Each readout is compared in terms of PSF and sampling efficiency.

*Readout trajectories*

All readout trajectories were based on rotations of an individual fundamental interleave:

**2D radial** – A full-spoke radial readout was rotated by a golden angle within each TR, covering a fully sampled disk during the whole acquisition schedule.

**2D spiral** – The fundamental waveform consisted of a variable density spiral interleave implemented using time-optimal gradient waveform design[31]. This waveform was rotated by a golden angle within each TR as in the 2D radial case.

**3D radial** – The same full-spoke radial readout used for 2D radial trajectory was rotated both around the z-axis (in-plane rotations) and the x-axis (spherical rotations) to cover a 3D volume of the *k*-space. A pseudo-random order of the spokes was achieved by random permutation of the readout order. The acquisition was performed using this ordering and using a single interleave within each TR.

**3D spiral** – The same variable-density spiral interleave used to construct 2D spiral trajectory was rotated along two different angles: in-plane rotations and spherical rotations. Full in-plane *k*-space coverage was achieved when the in-plane rotations equaled the number of interleaves in the waveform design. Spherical rotations enabled us to acquire data along all three spatial dimensions with fully sampled in-plane discs. Spherical rotations were incremented with the golden angle from one repetition to the next, while in-plane rotations were incremented linearly from one segment to the next.

*Acquisition and reconstruction pipeline*

All data were acquired following the excitation pattern shown in Fig. 1 with 880 repetitions per segment. Additionally, we accelerated the excitation pattern by reducing the total number of repetitions in Fig. 1a to 576, leading to a total acquisition time of 4:55 min. We acquired data with a

(192 mm)$^3$ field of view and reconstructed onto a 200$^3$ matrix, resulting in 0.96 mm$^3$ isotropic voxels. Table 1 summarizes the most important data acquisition parameters for all waveforms used in this work.

Acquired data were first anti-aliased with a *k*-space weighted view-sharing algorithm (Fig. 2) and reconstructed following the pipeline presented in Fig. 3. The view-shared data are first projected into a lower dimensional subspace with SVD compression to reduce the temporal signals to the first 10 SVD coefficients[32] and further reconstructed to obtain parametric maps and contrast-weighted images. In the following, we elaborate on steps b (*k*-space weighted view-sharing), f (parameter inference with artificial neural networks), and g (contrast-weighted image synthesis) of the acquisition and reconstruction pipeline (see Figure 3).

**Anti-aliasing via k-space weighted view-sharing**

To anti-alias prior to parameter inference, we used *k*-space weighted view-sharing. This algorithm builds upon the concept of *k*-space weighted image contrast[33] and has been used previously in the context of quantitative multiparametric mapping for Cartesian[19] and 2D radial imaging[20]. This method takes advantage of the fact that image contrast lies primarily within the central region of *k*-space, which is also oversampled. This enables the application of different filters in distinct *k*-space regions in order to enhance or reduce the contribution of each acquired view at each *k*-space location.

The method can be efficiently formulated as a linear weighted matrix multiplication in *k*-space, where the multiplication weights control the amount of sharing per view and *k*-space location. In our implementation, and different to Cruz et al.[20], we set the weights as a function of the sampling density in all three spatial dimensions instead of the distance to the center of *k*-space, enabling us to compare readouts with distinct density patterns (i.e. spirals and radials). A further explanation of our view-sharing algorithms can also be seen in Supplementary Figure 1.

In order to assess the contribution of view-sharing, we performed the following numerical experiment. We simulated the acquisition of time-domain data of a single representative white matter voxel (T1 = 600 ms, T2 = 70 ms) using all of the introduced undersampled trajectories. The data were then reconstructed with steps a-c of the pipeline to obtain a PSF in image space for each temporal coefficient. Data from the central 8 voxels were Fourier interpolated onto a 200-voxel grid for better visualization of the sidelobes. We reconstructed the data with both naïve zero-filling and view-sharing reconstruction. In addition, data corresponding to a fully sampled experiment were generated as a reference. We compared the results in terms of the suppression of the sidelobes and broadening of the PSF.

**Parameter inference with artificial neural networks**

Inspired by previous work[17], we rely on a fully-connected neural network for multi-parameter inference. By doing so, we embed the Bloch temporal dynamics by compact piecewise linear approximations rather than inefficient pointwise approximations used in the dictionary matching. In this work, we introduced the estimation of PD by developing and evaluating three distinct neural network architectures: a multi-pathway network with a single output parameter per pathway (*NN multipath*), a forward network with three output parameters (*NN fwd*), and an encoder-decoder network with two parameters in the encoded latent space and an output equivalent to the input (*NN fwd-bck*). This last alternative allows us to estimate PD in an identical manner to dictionary matching techniques.

*Neural network architectures*

The proposed neural network architectures are illustrated in Fig. 4. All models were designed to learn the non-linear relationship between the complex signal $x$ as input, and the underlying tissue parameters $\theta$, where the input layer of each network receives the complex signal in SVD space $x \in \mathbb{C}^{10}$ instead of the full temporal signal evolution. This lower dimensional input enables the network to be easily trained (e.g. using a CPU) and further avoids over-parametrization and the risk of overfitting. The input layer is followed by a phase alignment[17,34] to work with real-valued instead of complex-valued measurements, and a signal normalization layer.

Instead of a single-pathway architecture[17,35–37], our proposed *NN multipath* architecture (Fig. 4a) advances the use of multi-branch perceptrons for multitask problems[38], splitting parameter inference into individual pathways, wherein each specializes in the prediction of one parameter. Each branch comprises of 3 hidden layers with ReLU activations and 200, 100, and 1 node, respectively. The final nodes of these parametric pathways are then concatenated to form the parametric output $\widetilde{\boldsymbol{\theta}} = (\tilde{\theta}_1, \tilde{\theta}_2, \tilde{\theta}_3) \in \mathbb{R}^3$, representing an estimate for T1 and T2, along with the norm of the simulated signals as a proxy for PD. From this scaling factor, we can compute the relative PD $\tilde{\rho}$ as

$$\tilde{\rho} = \frac{\|x\|_2}{\tilde{\theta}_3}, \qquad [1]$$

where $\|x\|_2$ represents the 2-norm of the acquired signal $x$. The *NN fwd* architecture is a simplification of the *NN multipath* with the same total number of nodes, wherein all output parameters are inferred directly from the same hidden layers (Fig. 4b). Supplementary Figure 2 provides an additional comparison of the *NN multipath* against different single pathway architectures.

While both of these networks allow us to compute relative PD via Eq. 1, the mathematical formulation is not identical to dictionary matching techniques. Thus, we created a third architecture, *NN fwd-bck*, by concatenating a two-path network with a single-path model with reversed hidden layers (Fig. 4c). This architecture can be considered an autoencoder, as this model first infers parameters into a latent space and subsequently obtains an output signal $\tilde{x}$ equivalent to its input. In this model, the encoding portion is used to infer T1 and T2 while the decoding portion acts as a Bloch simulator to create the output signal $\tilde{x}$. It is then possible to infer the relative PD $\tilde{\rho}$ in an identical way as is done with dictionary matching techniques

$$\tilde{\rho} = \frac{\langle x, \tilde{x} \rangle}{\|\tilde{x}\|_2}. \qquad [2]$$

In all formulations, the inference of the relative PD within the neural network framework allowed us to use it along other quantitative parameters to subsequently synthesize multiple clinical contrasts via the solutions of the Bloch equations. Figure 5 illustrates the data processing pipeline from a temporal viewpoint, including neural network parameter inference (Fig. 5f) and contrast synthesis (Fig. 5g).

*Neural network training and validation*

Based on Extended Phase Graphs[39], we created a dataset of synthetic signals of 52,670 samples for T1 = [10 ms, 5000 ms] and T2 = [10 ms, 2000 ms] to train our neural networks and to obtain a dictionary matching reference. T1 values were simulated in steps of 1 ms from 10 ms to 2000 ms and in steps of 100 ms from 2100 ms to 5000 ms. T2 values were simulated in steps of 5 ms between 10 ms and 300 ms, and in steps of 10 ms between 310 ms and 2000 ms. In fact, we trained each neural network individually for each set of acquisition parameters (2D and 3D spirals and radials). This was done because the 3D scheme varies slightly from the 2D implementation, as the 3D acquisition requires a second iteration for the signal (Figure 1b) and we included the slice profile in the simulation[40,41] for the 2D case. Also, we used a different TR for 1.5T and 3T acquisitions.

For robust model training, we corrupted 80% of the generated samples with zero-mean independent Gaussian noise and randomly varied SNR between 30 to 70 dB. We trained all networks on input batches of size 100, using gradient descent optimization with a learning rate of 0.005 and the loss function $J$ defined as the mean absolute percentage error

$$J = \frac{1}{N} \sum_{n=1}^{N} \left| \frac{\boldsymbol{\theta'}^n - \widetilde{\boldsymbol{\theta'}}^n}{\boldsymbol{\theta'}^n} \right|, \qquad [3]$$

where n ranges over all samples in the training set, $\widetilde{\boldsymbol{\theta'}}^n$ is the multiparametric output vector, and $\boldsymbol{\theta'}^n$ the corresponding ground truth for sample $n$. The dropout rate was set to 0.9 and the model was trained for 500 epochs, keeping the model that achieved the best performance on the validation data, i.e. the remaining 20% of the samples. The stopping criterion was found empirically from previous experiments as the validation loss was stable at this point. All networks were developed using TensorFlow.

To evaluate network inference we compared the reconstruction quality and computation times with exhaustive grid search (dictionary matching) and fast group matching[42], which was implemented with 80 groups and SVD compression[32]. Image reconstruction and parameter inference were performed on an Intel Xeon processor E5-2600 v4 (48 CPU cores) equipped with a NVIDIA Tesla K80 GPU (using the gpuNUFFT package for regridding[43]).

**Contrast-weighted image synthesis**

To allow radiological assessments, we used the parameters inferred from the neural network to synthesize three different clinical contrasts: (1) a spoiled gradient recalled echo (SPGR) with TR = 5.83 ms, FA = 13°, (2) a fast spin echo (FSE) using TE = 100 ms; and (3) a fluid attenuated inversion recovery (FLAIR) based TE = 84.812 ms, and TI = 2500 ms. In SPGR, contrast is dominated exclusively by TR and T1

$$S_{GRE} = \rho \frac{\sin \alpha \, (1 - e^{-TR/T1})}{(1 - \cos \alpha \, e^{-\frac{TR}{T1}})}, \quad [4]$$

while FSE contrast is driven by TE and T2

$$S_{FSE} = \rho e^{-TE/T2}, \quad [5]$$

and FLAIR contrast is determined by both T1 and T2:

$$S_{FLAIR} = \rho e^{-TE/T2}(1 - 2e^{-TI/T1}). \quad [6]$$

To evaluate the effectiveness of our synthetic images, we compared them with standard acquisitions with matching imaging parameters.

*In vitro* **validation**

We acquired the same tubes of the Eurospin TO5 phantom[44] on a 1.5T (GE HDxt) and 3T (GE 750w) using all waveforms. To evaluate the impact of view-sharing, we reconstructed the data using either only SVD projection[32] or *k*-space weighted view-sharing followed by SVD projection (KW-SVD), and subsequently inferred the parameters via each trained neural network. To assess accuracy and precision of the predictions, bias and standard deviation of T1/T2 values were calculated for each vial. In addition, concordance correlation coefficients[45] (CCC) and normalized root mean square errors (NRMSE) were calculated at both field strengths for each trajectory and for both SVD and KW-SVD reconstructions. NRMSE was normalized with respect to the average of the nominal values of the vials. As a reference, we used the gold standard values reported in the phantom's manual together with our own measurements using FISP-MRF from Jiang *et al*[46], as it has been previously validated for repeatability and reproducibility[47,48].

*In vivo* validation

The study was conducted according to the principles of the Helsinki Declaration. All subjects gave their informed consent to the enrollment in the study protocol GR-2016-02361693, approved by the local competent ethical committee, the Comitato Etico Pediatrico Regionale (CEPR) of Regione Toscana, Italy. We used the same protocol to acquire volunteer data at 1.5T and 3T, reconstructing the data with and without view-sharing and estimating parameters with dictionary matching and neural network inference. Representative ROIs were manually drawn in white matter and gray matter to quantitatively compare our results to previous literature reports.

# Results

**Impact of readout trajectories and view-sharing reconstruction**

Figure 6 shows the results of the PSF analysis. Our PSF analysis shows that all undersampled readouts approximate the fully-sampled PSF for the first SVD coefficient, whereas view-sharing significantly improves the PSF with respect to zero-filling in other lower energy coefficients, such as the ninth, where the sidelobes of the PSF were reduced by an order of magnitude.

Figure 7 showcases the results for 1.5T, where we focus on contrasting the different readout trajectories and the impact of SVD compression alone versus view-sharing followed by SVD compression (KW-SVD). The benefits of view-sharing can be most clearly visualized in the T1 maps for 3D radial and 3D spiral trajectories, where higher spatial consistency can be achieved without negatively impacting sharpness or quantification accuracy. However, trajectories with high undersampling factors, such 3D radial, still present clearly visible artifacts that translate into blurred T2 maps. Overall, 3D spirals achieved the most efficient sampling coverage and in combination with view-sharing, could best approximate the fully sampled PSF throughout different spatiotemporal coordinates.

**Neural network inference**

Figure 8 compares dictionary matching against multi-path neural network inference for the *in vitro* validation using the 3D spiral acquisition at 3T. Here, we observe a high degree of correlation between dictionary matching and network inference, where 95% confidence intervals lie within 6.58% difference of each other. We further validated the neural network inference results against gold standard phantom references for both the standard case and the accelerated, high-resolution version. We display these results in Supplementary Figures 3-5 and Tables 2 and 3, and elaborate on them in the next section.

Trained once, neural networks allow us to infer high resolution maps with quantification accuracy and image quality compared to dictionary matching results as seen in Fig. 9, which presents representative 3T results based on the 3D spiral acquisition scheme. The high degree of agreement between neural network inference and dictionary matching is confirmed by Fig. 10, where we depict a voxel-wise comparison based on the percentage difference between neural network prediction and dictionary matching results.

Here, we specifically observe that PD estimates obtained with direct inference with the three-pathway network *NN multipath* (via Eq. [1]) and with the autoencoder network *NN fwd-bck* (via Eq. [2]) are more consistent with dictionary matching than the results of the single pathway network *NN fwd*. For the latter we visually observe diminished tissue contrast in PD and higher percentage differences also for T1 and T2. We attribute these differences to the simplicity of the *NN fwd* architecture, evidencing how naïve single-path architecture with shared weights can result in declined performance for a subset of the output parameters. A multi-path architecture on the other hand gives more flexibility to learn the physical relationship between the input signal and the individual parameter. That is, the separate pathways of *NN multipath* and *NN fwd-bck* ensure that features that are essential for the recovery of the individual parameters are retained, and not suppressed by other features which is a risk a single-path schemes with shared weights. Thus, the *NN multipath* and *NN fwd-bck* produce higher fidelity results as compared to a dictionary, and their modular architecture potentially enables the inclusion of additional parameters beyond the ones considered in this study.

With the estimation of PD as an integral part of the neural network inference, it is now possible to produce synthetic images from the inferred parameters showing typical contrasts used during radiological evaluations. Figure 11 compares three typical contrasts – GRE, FSE, and FLAIR – obtained with our proposed technique against standard clinical protocols. Moreover, Supplementary Figure 6 displays a visualization of each of these synthetic contrasts along the axial, sagittal, and coronal planes.

**Quantification accuracy and precision**

T1 and T2 quantification results with neural network inference as compared to the gold standard reference are summarized in Supplementary Figures 3-5 alongside Tables 2 and 3. Results for both 1.5T and 3T show that acquisitions using all tested readout trajectories are comparable to the reference and could be used for multiparametric mapping within clinical settings: maximum quantification bias for T1/T2 was less than 9/13% at 1.5T and less than 9/15% at 3T (Supplementary Figure 4), while CCC were higher than 0.98 for both T1 and T2 at 1.5T, 0.92 for T2 quantification with the 2D spiral readout at 3T, and above 0.97 for all other cases (see Table 3).

While all the trajectories achieved a similar accuracy, 3D spiral trajectories showed substantially reduced undersampling artifacts, resulting in a diminished standard deviation of the measurements (3-15% for T1, 4-11% for T2 at 1.5T; 3-14% for T1, 4-10% for T2 at 3T). Conversely, quantification errors due to aliasing errors for highly undersampled trajectories (such as 3D radial) were still present, leading to high standard deviations (6-22% for T1, 11-27% for T2 at 1.5T; 5-18% for T1, 10-29% for T2 at 3T), as shown in Supplementary Figure 4.

All trajectories (especially 3D trajectories) benefitted from the KW view-sharing processing, resulting in a substantial reduction of aliasing artifacts: at 1.5T, standard deviation contracted from 3-26% to 3-16% for T1 and from 4-27% to 3-18% for T2, while at 3T it diminished from 3-19% to 2-13% for T1 and from 4-29% to 3-20% for T2 (Supplementary Figure 4). In addition, KW view-sharing added only a minimal impact on the quantification bias. The net effect was a reduction of the NRMSE of most of the measurements (see Table 2): at 1.5T, NRMSE decreased from 5.6-9.4% to 4.2-6.6% for T1 quantification of 3D readout trajectories, while it increased slightly from 8.4-10.26% to 8.5%-11% for 2D trajectories. For T2, it reduced from 7.0-15.3% to 4.9-11.1% for all trajectories. At 3T the reduction was again observable in 3D trajectories for T1 from 7.7-9.5% to 6.4-8.7% and in all trajectories for T2, from 9.6-13.5% to 7.6-10.1%. We observed an increase in the NRMSE for T1 quantification for 2D radial and 2D spiral measurements at 3T from 6.2-8.3% to 10.1-10.4%.

In vivo experiments confirmed the results of the phantom experiment: T1/T2 values resulted in 600-774 ms / 53-66 ms for white matter (compared to 788-898 ms / 63-80 ms reported in literature[49–51]), and 1004-1381 ms / 89-132 ms for gray matter (previous reports are 1269-1393 ms / 78-117 ms). Reduction in standard deviation due to KW view-sharing was comparable to the phantom experiment.

**Acquisition and reconstruction efficiency**

Table 1 reports acquisition times for all experiments, whereas Table 4 compares reconstruction and inference times of dictionary matching, group matching, and neural network parameter inference. For all cases, neural network inference significantly outperformed dictionary matching and fast group matching techniques in terms of computation efficiency. In the accelerated 3D experiment, we achieved submillimeter resolution with an acquisition time of 4:55 minutes and reconstruction and parameter inference times of 6:07 minutes using the proposed view-sharing technique and GPU gridding. The results of this high-resolution and accelerated mapping experiment are displayed in Fig. 12. For a clearer visualization of the brain, we removed all non-brain tissue with the FSL brain extraction tool[52].

## Discussion

In this work, we developed a quantitative parameter inference pipeline based non-iterative anti-aliasing, which takes advantage of the computational efficiencies arising from subspace dimensionality reduction and neural network parameter inference. We compared the performance of this pipeline on data obtained from different 2D and 3D non-Cartesian trajectories. Our systematic comparisons confirmed that that spiral *k*-space trajectories have higher sampling efficiency than radial. On the other hand, all readouts benefitted from *k*-space weighted view-sharing.

All waveforms studied had a higher sampling density near the center of *k*-space, thus could accurately capture contrast information while relying on view-sharing to recover the sparsely sampled outer portions of *k*-space. Importantly, we showed that quantification values of T1 and T2 did not depend on the specific trajectory used and agreed with the reference values from the Eurospin TO5 datasheet. On the other hand, slight T2 underestimation was observed for all our measurements *in vivo*, when compared to literature value (steady-state or spin echo method). We hypothesize this is due to effects not included in our inference model, such as unmodelled *k*-space trajectory errors, diffusion or magnetization transfer effects. Further extension of the physical models at the basis of our inference can establish the impact of these effects on accuracy and precision of the estimates.

A general limitation of performing inference on undersampled transient-state imaging is the presence of errors due to aliasing, as a consequence of undersampling the *k*-space differently in each repetition. Artifacts in image domain can be reduced by temporal compression, but temporal compression models usually do not take *k*-space encoding into account, and specific anti-aliasing strategies have proven advantageous prior to estimating the parameters[4,21,22,53]. Amongst anti-aliasing methods, iterative approaches have limitations for full three-dimensional non-Cartesian acquisitions, where 3D gridding and inverse gridding would be required at each step, significantly increasing the computational burden associated with spatial decoding. We found that local quantification errors due to aliasing can be reduced by a simple non-iterative anti-aliasing technique, allowing clearer images for radiological evaluations. The concept used here for anti-aliasing is similar to keyhole imaging[20,33], and assumes that the image contrast is mainly stored in the center of *k*-space, while the image details, which are mostly unchanged between frames, are in the edges of *k*-space. As in principle the effect of such temporal filters on the final map accuracy is unknown, the impact of *k*-space weighted view-sharing on the PSF was systematically evaluated in the SVD space, finding that it has very little impact on the blurring of the images, but greatly reduces the local quantification errors due to aliasing. As all the operations are linear, the process can be formulated as a series of simple operations, having a minor impact on reconstruction time.

We have demonstrated that anti-aliasing techniques used reduce artifacts on the final maps, while introducing only minimal apodization. However, our method still relies on assumptions on the *k*-space distribution of contrast information, and highly undersampled acquisitions can still present visible artifacts, like for the case of the 3D radial trajectory shown here. Methods tackling the optimization directly in the time domain, such as MR-STAT[5], have a superior artifact robustness because aliasing is not present in *k*-space. However, this comes at the expense of computation time. Times reported for multi-slice MR-STAT have been of the order of several hours on scientific computing clusters, which can limit clinical applicability. Future improvement of methods including full spatial modelling such as MR-STAT could potentially improve on the efficiency of the estimates presented here while keeping reasonable reconstruction times.

An important aspect of the method described here is the relatively high efficiency at encoding the tissue properties into the transient-state signals. While here we have used a sequence with a simplified flip angle schedule, various different strategies to optimize the acquisition are possible. For instance, it is possible to optimize the Cramér-Rao Lower Bound of quantitative sequences[9,1314], recently demonstrated also in combination with automatic differentiation algorithms, hence without approximations or an analytical formulation[9]. It is also possible to use Bayesian design theory to define a set of optimal acquisition parameters for a particular range of tissues of interest, maximizing both parameter encoding and experimental efficiency[4]. This approach could also be used to systematically encode other MR-sensitive parameters such as diffusion or magnetization transfer into transient-state signals. In alternative to transient-state acquisitions, pseudo-steady state methods (pSSFP)[54,55] have demonstrated improved efficiency in terms of parameter encoding capabilities and the use of pSSFP may further improve on the results shown here.

As an alternative to approaches using an exhaustive search over a grid representing a dictionary of simulated evolution, performing the parameter estimation step via a neural network has several advantages[28]. First, storage requirements for the network are much less restrictive than those of the dictionary. Second, parameter estimations using neural networks are significantly faster than exhaustive searches over a predefined grid, especially if the model is estimated on a GPU. Third, under the assumption of local linearity of the Bloch manifold, the network is not limited to a discretized set of parameter values. Our proposed network architectures were inspired by recent work at the intersection of deep learning and quantitative mapping techniques, such as the MRF-Net proposed in the study on geometry of deep learning for MRF[17] and the deep reconstruction network (DRONE)[35]. Both works demonstrated that dictionary matching for T1 and T2 mapping can be replaced with high accuracy by a compact, fully-connected neural network trained on simulated signals. Moreover, whereas the DRONE requires the full temporal signal and thus is computationally prohibitive for high-resolution 3D mapping, the MRF-Net performs inference on dimensionality reduced signals. Our networks are motivated by these works, whereas the proposed multi-pathway and autoencoding architectures alongside the integrated PD estimation are a significant technical novelty with respect to previous implementations.

Only now, the neural network-based parameter inference can fully replace the dictionary matching framework as all previous works lacked the inference of a PD estimate which is required for subsequent synthetic image generation. Also, the proposed multi-pathway model design presents an opportunity for compact transfer-learning. Previous works either focused on single-pathway implementations where all output parameters emerge from the same fully-connected branch or trained individual networks for each parameter to be estimated[37,56,57]. Here, we formulate the multivariate regression as a single joint optimization, but give the network more freedom in finding the non-linear input-output mappings compared to single-pathway schemes, as each pathway can specialize in the inference of one parameter. At the same time, we prevent overfitting of one specific task due to the regularizing and balancing capacity of the joint loss. The increased flexibility provided by the multi-pathway model could also facilitate an extension to additional parameters as MR sequences are continuously developed to simultaneously encode for more information.

Other recent works rely on spatially constrained, convolutional neural networks, such as U-Net architectures[37,56], for parameter inference in high-resolution MRF experiments. Although these approaches have shown to provide high quality parameter maps, they are based on the dictionary matching results for model training, which cannot be considered a ground truth. In contrast to the recently proposed U-Net architectures, we present a compact model that facilitates training and parameter inference even on CPUs. Our neural network is trained on purely synthetic signal time-courses. Its performance is independent of and hence not bound by the quality and accuracy of dictionary matching results. Once trained, the proposed model is then universally – in terms of anatomical region and patient pathologies – applicable in a clinical setting. This might not be the case for spatially constrained networks which have only seen e.g. healthy brain tissue during training.

The combination of simple and efficient anti-aliasing with a multi-pathway neural network parameter inference allowed us to reconstruct a full 3D volume with sub-millimeter resolution in under 7 minutes using a high-performing server (see system details in Table 4). We consider this an important result, as current state-of-the-art achieves comparable reconstructions within 4 hours (see Table 5) and is limited by the dictionary discretization grid[11].

## Conclusion

Three-dimensional quantitative transient-state imaging provides an efficient framework for multiparametric mapping, achieving high anatomical detail and quantification accuracy and precision. The novel introduction of proton density in a compact, multi-path neural network, coupled with a computationally rapid anti-aliasing technique, allowed us to acquire and reconstruct full volume, high-resolution isotropic data within clinically acceptable times. Given this fast and comprehensive acquisition, the method proposed here can be used in challenging populations, including elderly patients and children. Our improved acquisition and reconstruction times encourage further investigation towards clinical applications.

## Data availability

The datasets generated during and/or analyzed during the current study are available from the corresponding author on reasonable request.

## Acknowledgements

This research was funded by the Italian Ministry of Health and by Tuscany Region under the project 'Ricerca Finalizzata', grant no. GR-2016-02361693. Funding from the EMPIR Programme (18HLT05 QUIERO Project) co-financed by the Participating States and from the European Union's Horizon 2020 Research and Innovation Programme. CP acknowledges Deutsche Forschungsgemeinschaft (DFG) through TUM International Graduate School of Science and Engineering (IGSSE), GSC 81.

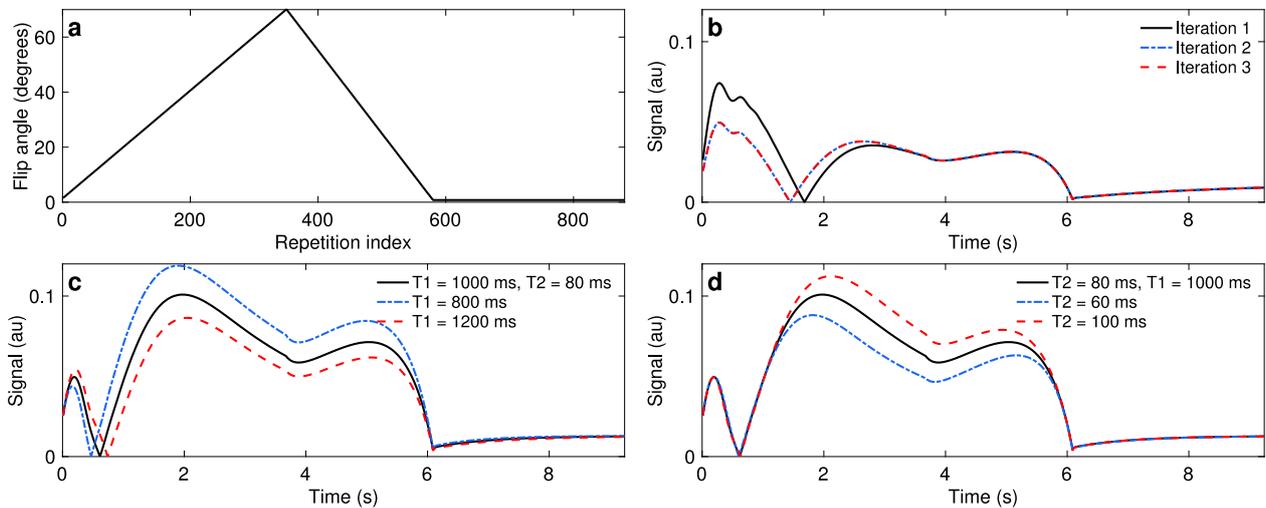

*Figure 1: Excitation pattern. **a,** The excitation pattern consists of a flip angle ramp to encode for T1 and T2 followed by constant, small flip angles to allow for T1 recovery before the next inversion pulse. **b**, Signal evolutions reach a 'steady' transient-state after two iterations as magnetization before the inversion exhibits the same initial conditions after the first iteration. **c-d,** T1 and T2 encoding: T1 is encoded into the signal through the inversion pulse prior to the flip angle train (**c**), whereas larger flip angle values produce stimulated echoes for T2 encoding (**d**).*

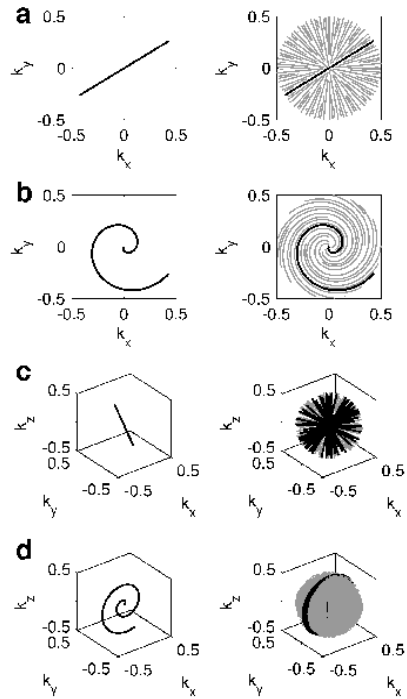

*Figure 2: k-space weighted view-sharing with different acquisition trajectories.* k-space weighted view-sharing consists of sharing spatial acquisition data within neighboring temporal frames to increase the number of samples per frame. This technique can be applied to all non-Cartesian trajectories, i.e. 2D radial (*a*), 2D spiral (*b*), 3D radial (*c*) and 3D spiral (*d*) readouts.

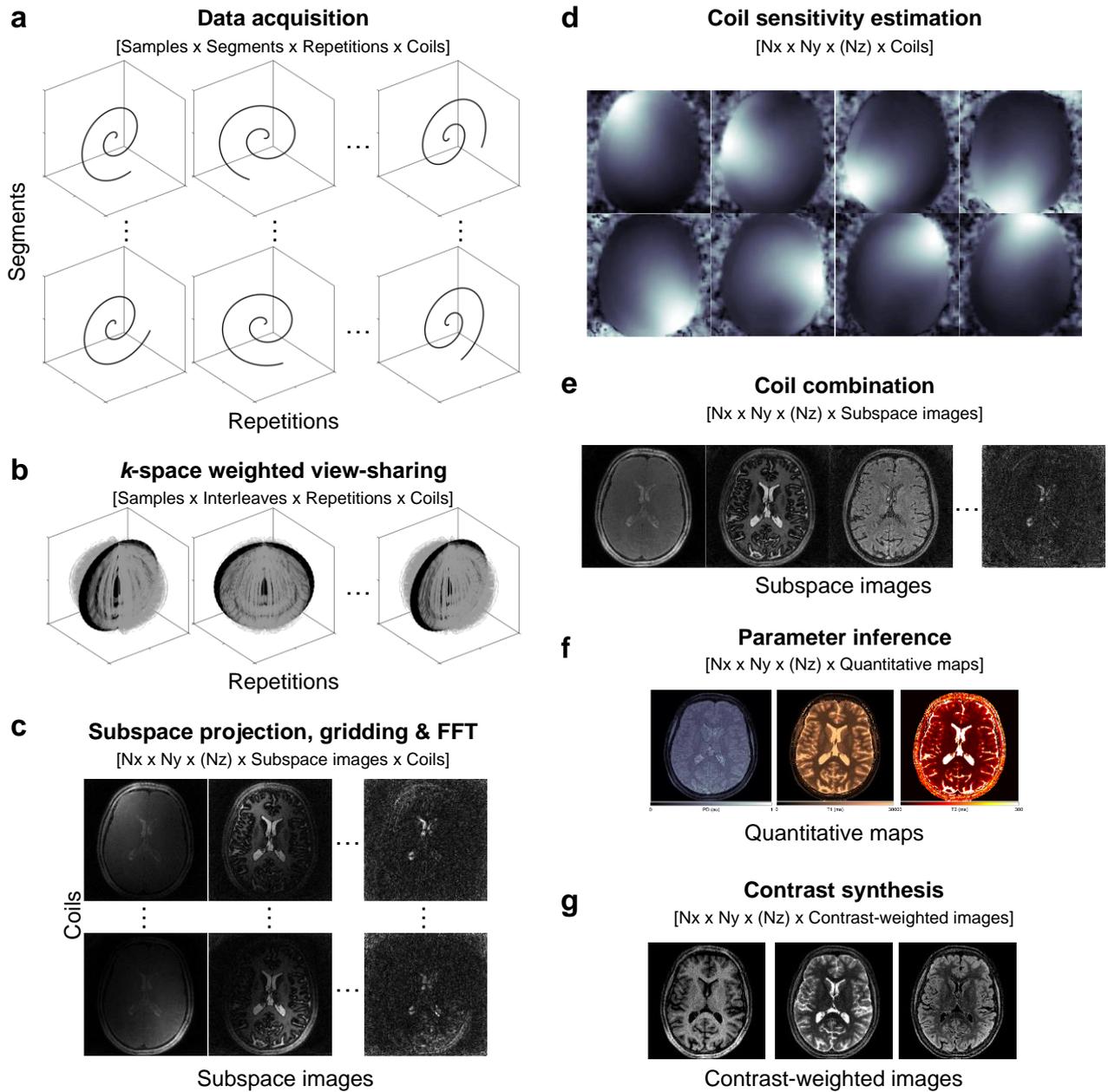

*Figure 3: Data acquisition and reconstruction pipeline. **a**, Undersampled data are acquired throughout different repetitions with unique contrasts and distinct segments with equivalent contrasts. **b**, k-space weighted view-sharing increases the total amount of acquired samples per repetition. **c**, Data are dimensionality reduced and reconstructed to produce one image per coil and subspace coefficient. **d-e**, Sensitivity maps are computed and data from different receiver coils are combined. **f**, Parameter inference produces quantitative maps as an output. **g**, The maps can subsequently be used to synthesize contrasts for qualitative imaging.*

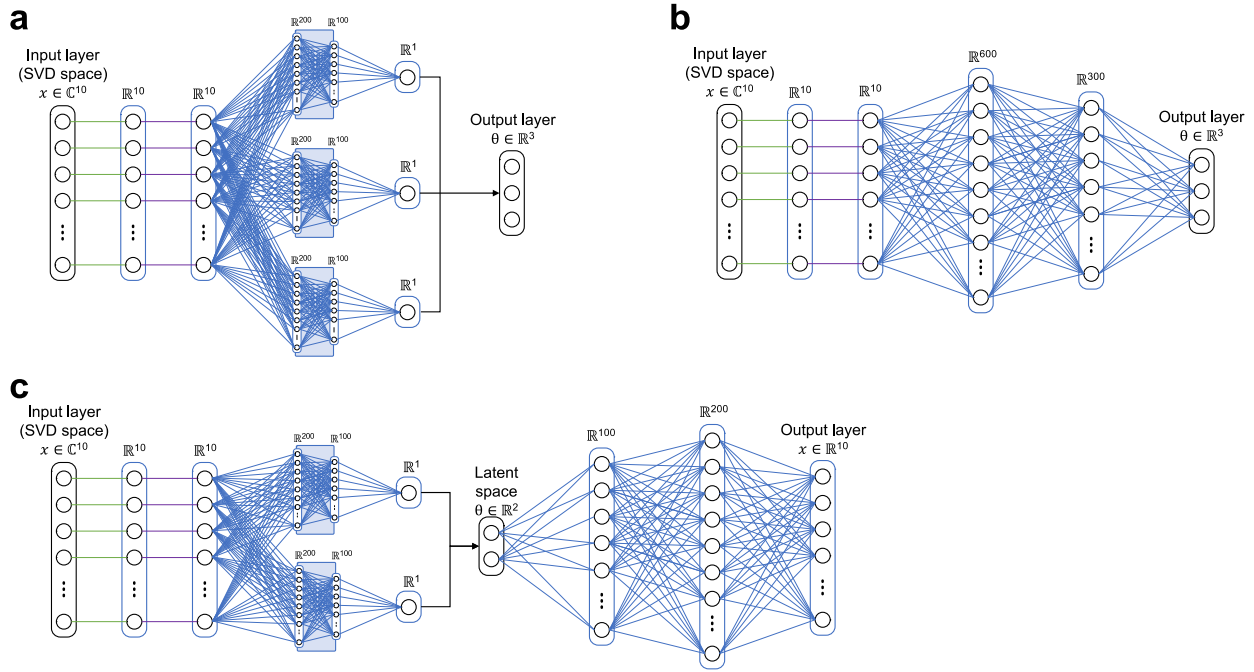

*Figure 4: **Neural network architectures for parameter inference.*** *The neural networks receive the complex signal in SVD subspace and infer the underlying tissue parameter vector **θ** based on different architectures. **a**, The multi-pathway implementation (NN multipath) is designed such that each path specializes on the estimation of one parameter, parameter $\theta_i$, i.e. T1, T2 or a PD related scaling factor. **b**, The single pathway (NN fwd) architecture has the total number of nodes as the NN multipath but directly infers the parameter vector from the same hidden layers. **c**, Parameters can also be derived from an autoencoder network (NN fwd-bck), where T1 and T2 are directly inferred from the latent space and relative PD is computed from the decoded output signal.*

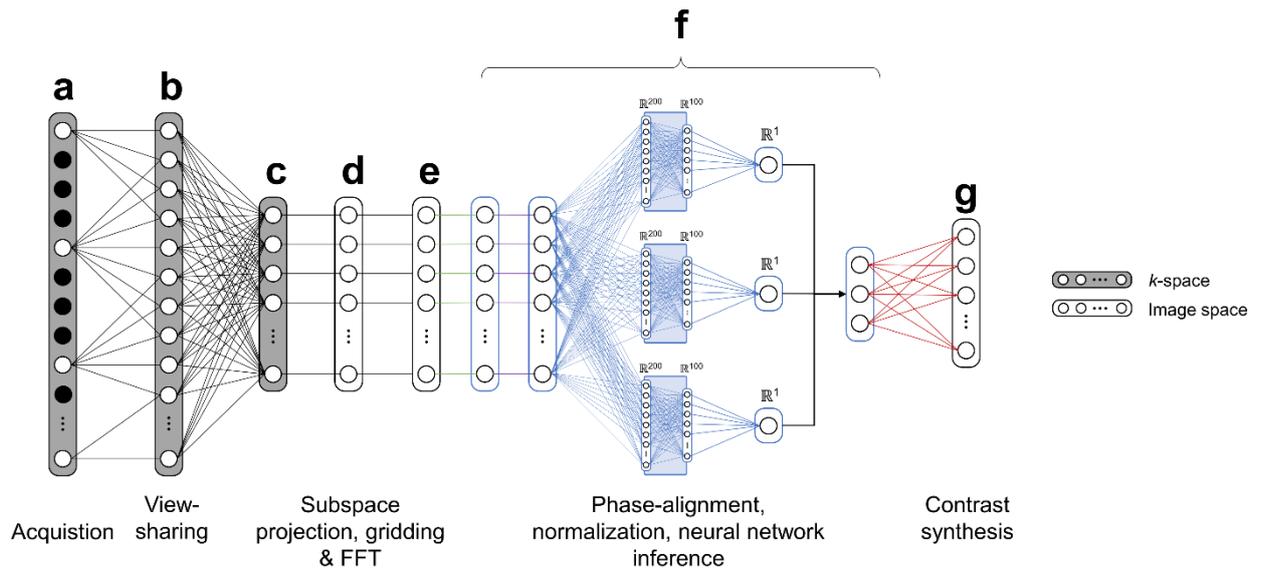

*Figure 5: Temporal data processing pipeline with neural network parameter inference. The letters follow the same order as Fig. 3, where the neural network architecture corresponding to the multi-pathway network is shown in **f**. After dimensionality reduction via SVD subspace projection (**c**), gridding, FFT, coil estimation and combination (**c-e**) are all linear operations that do not affect the input of the compressed signal into the neural network. The output of the network is subsequently used to synthesize image contrasts (**g**).*

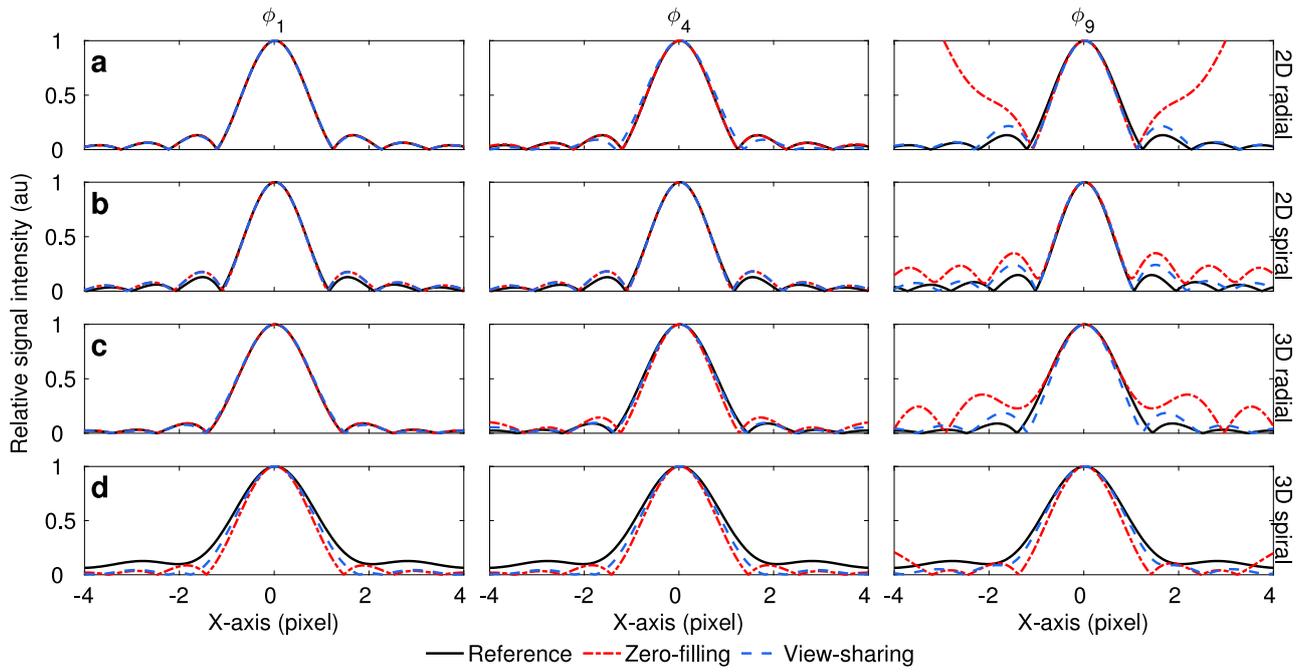

*Figure 6: **Point spread function analysis.*** *The PSF analysis evidences that view-sharing reconstruction is beneficial when considering lower energy SVD coefficients. Without view-sharing, the PSF will have increased sidelobes for all trajectories, most notably in 2D and 3D radials.*

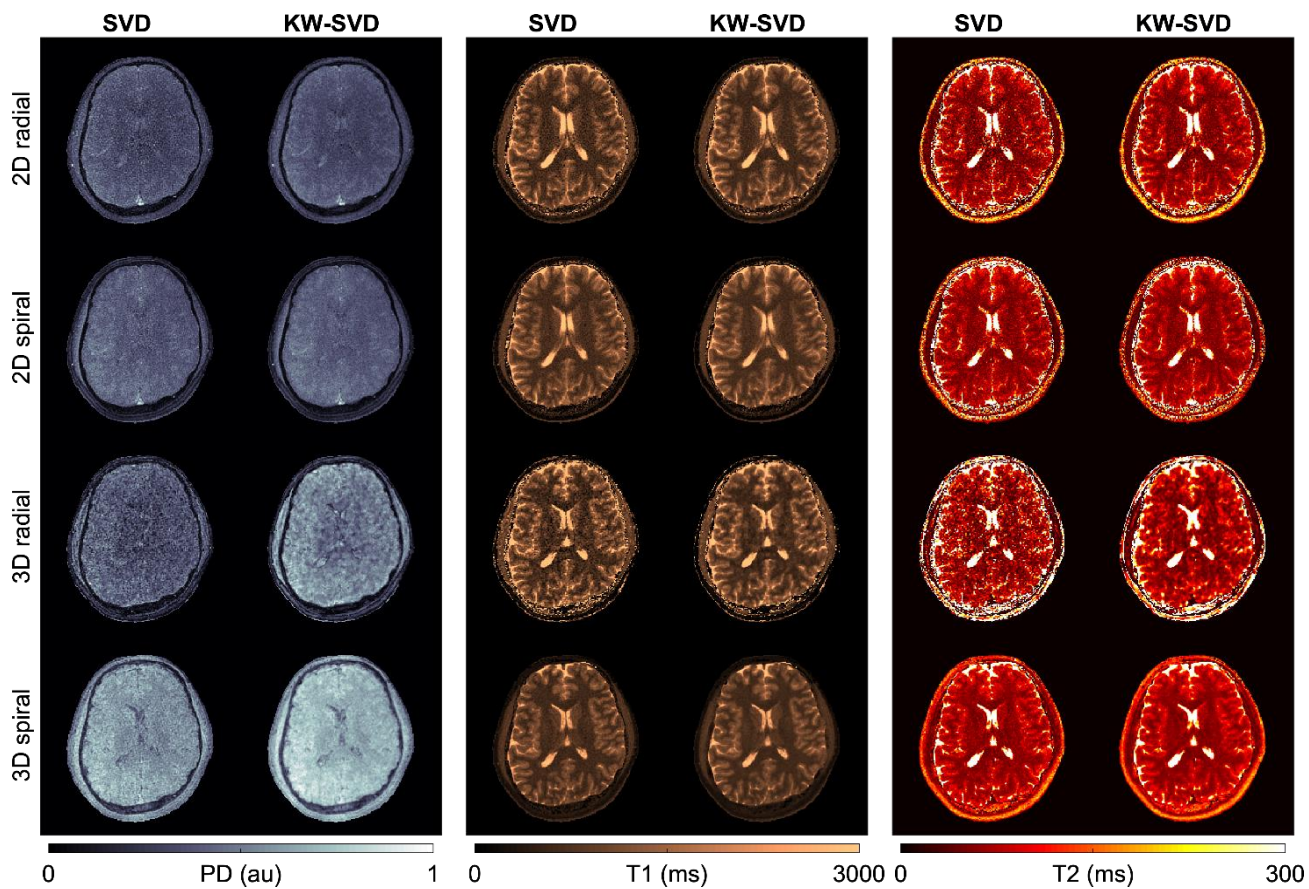

*Figure 7: Readout trajectories and reconstruction comparison.* *A comparison of the different readout trajectories evidences that, while all could be used multiparametric mapping, 3D radials suffer from larger undersampling artifacts. Moreover, view-sharing improves spatial consistency without impacting parameter quantification.*

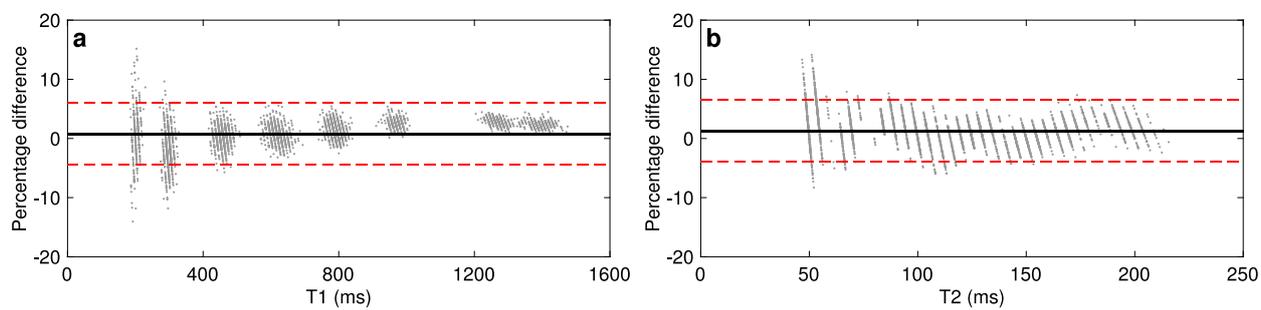

*Figure 8: Multipath neural network-based parameter inference – in vitro validation.* 95% confidence intervals in percentage differences to the dictionary are -4.45% – 6.00% for T1 and -3.97% – 6.58% for T2.

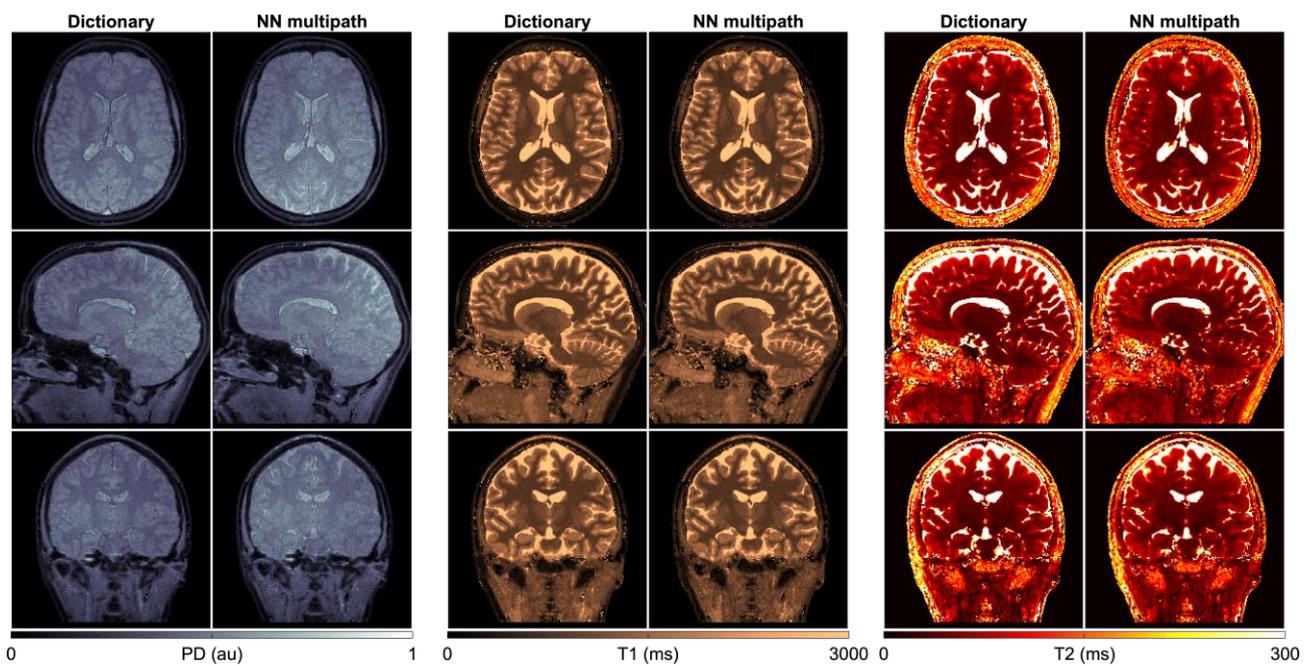

*Figure 9: Parameter inference via multipath neural network inference and dictionary matching.* Inferring parameter maps with a trained neural network and a high-resolution dictionary produces comparable results – with the key difference that the network is not limited by dictionary size or its discretization grid, and outperforms exhaustive grid search in terms of required memory and computation times.

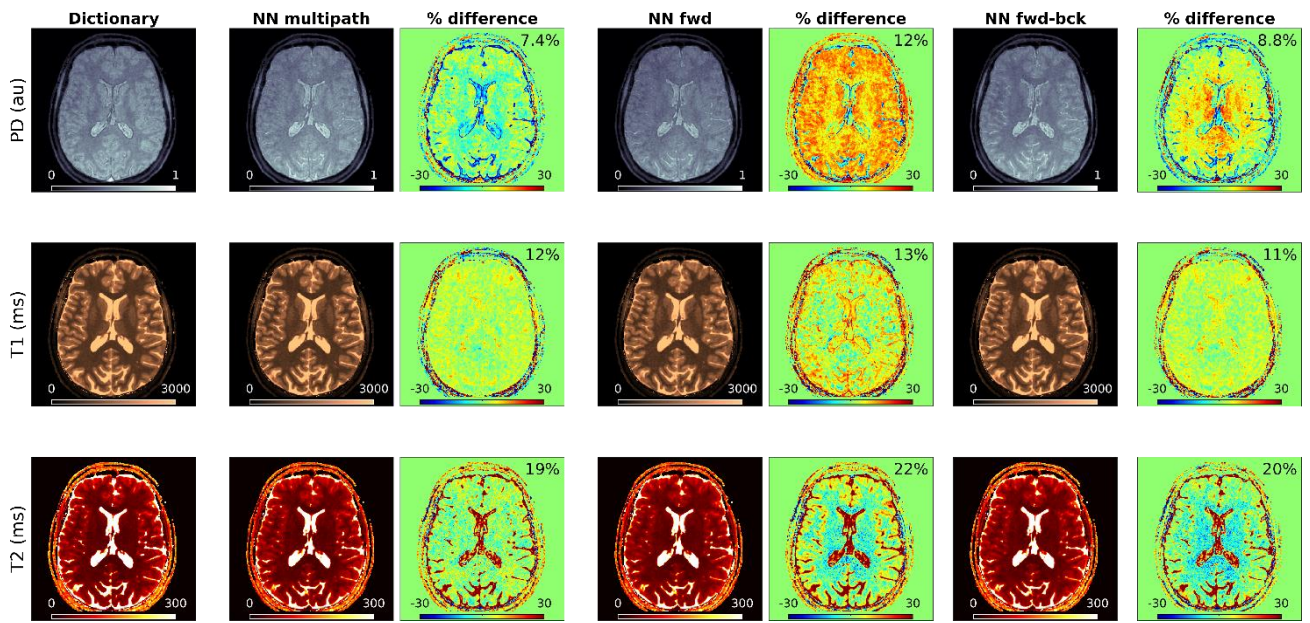

*Figure 10: Neural network-based parameter inference – in vivo validation.* All neural network implementations provide parametric estimates that are largely consistent with dictionary matching results. For all parameters, the neural network architectures with separate pathways, i.e. NN multipath and NN fwd-bck, resulted in closer estimates with respect to the dictionary than NN fwd, which is based on a single-pathway scheme.

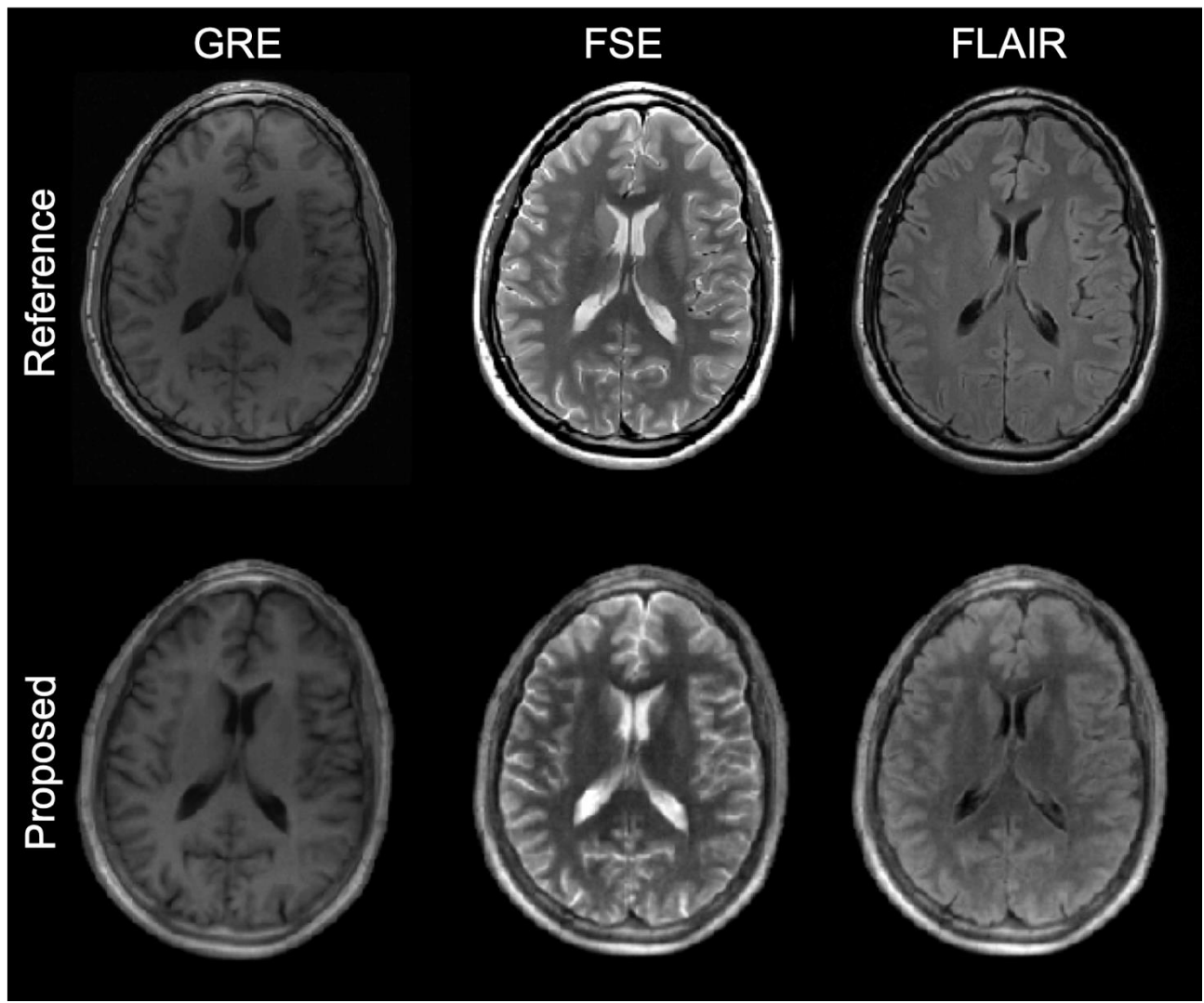

*Figure 11: Contrast-synthesis.* *The inclusion of PD to the inferred parameters enables high-quality contrast-synthesis.*

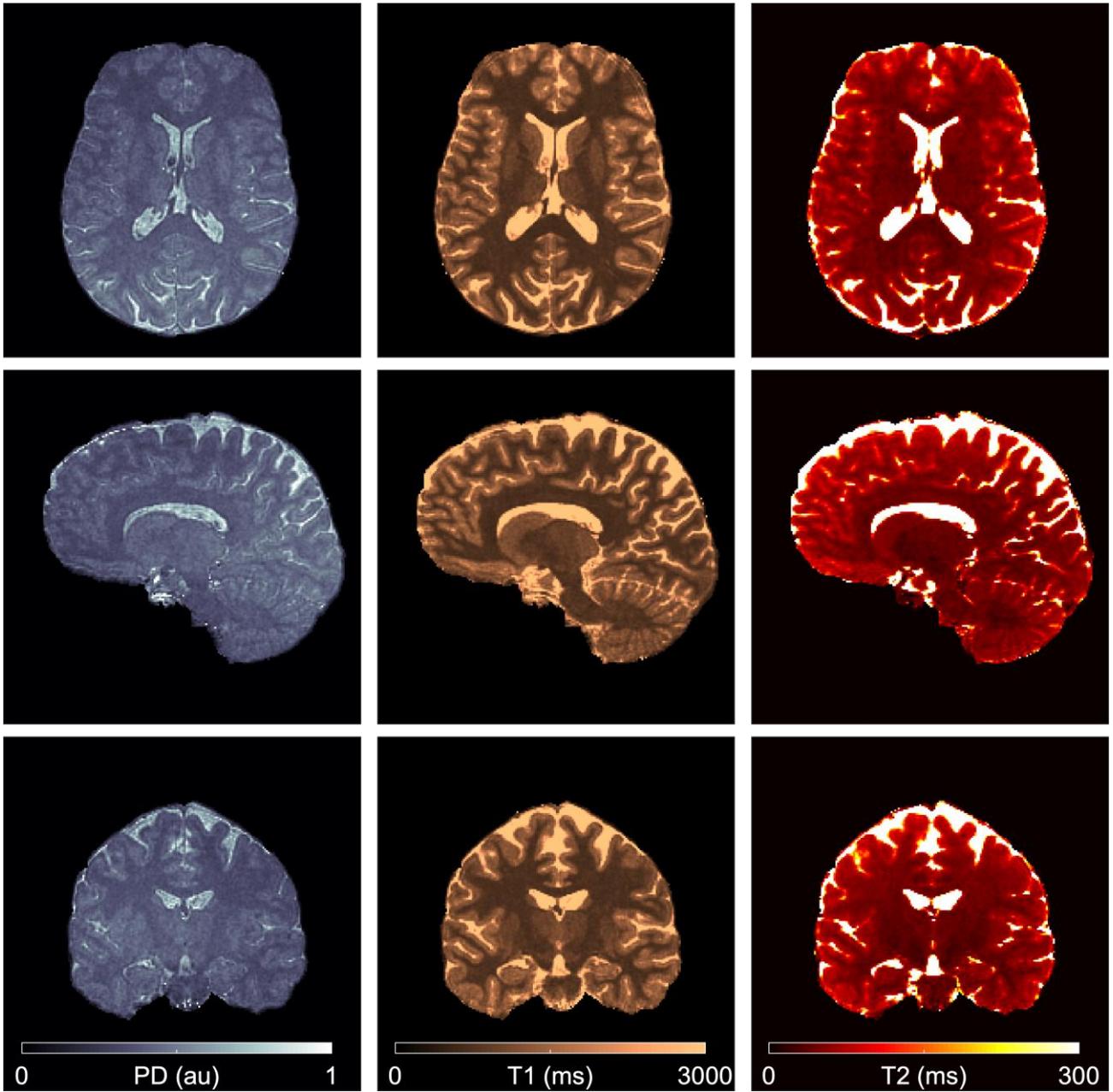

*Figure 12: Accelerated mapping.* The method presented here achieves submillimeter isotropic resolution multiparametric mapping in under 5 minutes with efficient reconstruction and neural network parameter inference in under 7 minutes.

*Table 1: Readout trajectories and experimental details.*

| Parameter | 2D radial | 2D spiral | 3D radial | 3D spiral | 3D spiral (high-res) |
|---|---|---|---|---|---|
| Gradient amplitude (mT/m) | 10 | 16 | 10 | 18 | 10 |
| Slew rate (T/m/s) | 70 | 80 | 70 | 60 | 90 |
| Field of view (mm) | 200 | 200 | 200 | 200 | 192 |
| Matrix (mm) | 200 | 200 | 200 | 200 | 200 |
| Resolution ($mm^2/mm^3$) | 1 | 1 | 1 | 1 | 0.96 |
| Waveform duration (ms) | 4.15 | 4.91 | 4.15 | 4.85 | 6.88 |
| Acquired *k*-space samples | 496 | 920 | 496 | 876 | 1348 |
| Sampling time (ms) | 1.98 | 3.68 | 1.98 | 3.50 | 5.39 |
| TE (ms) | 2.08 | 0.46 | 2.08 | 0.46 | 0.46 |
| Number of in-plane interleaves | 987 | 377 | - | 55 | 48 |
| Number of spherical rotations | - | - | - | 880 | 576 |
| Total interleaves | 987 | 377 | 63160 | 48400 | 27648 |
| Acquired interleaves | 880 | 880 | 49280 | 49280 | 28224 |
| Acquisition time @1.5T | 10.56 s | 10.56 s | 9:51 min | 9:51 min | - |
| Acquisition time @3T | 9.24 s | 9.24 s | 8:37 min | 8:37 min | 4:55 min |

*Table 2: Phantom validation results for parameter inference using the NN multipath architecture.* NRMSE normalization were calculated with respect to the average nominal value of the vials.

| | | T1 1.5T | | |
|---|---|---|---|---|
| | CCC SVD | CCC KW-SVD | NRMSE SVD [%] | NRMSE KW-SVD [%] |
| 2D radial | 0.99 | 0.98 | 10.26 | 11.0 |
| 2D spiral | 0.99 | 0.99 | 8.4 | 8.5 |
| 3D radial | 0.99 | 0.99 | 9.4 | 6.6 |
| 3D spiral | 0.99 | 0.99 | 5.6 | 4.2 |
| | | T2 1.5T | | |
| | CCC SVD | CCC KW-SVD | NRMSE SVD [%] | NRMSE KW-SVD [%] |
| 2D radial | 0.99 | 0.99 | 14.7 | 11.1 |
| 2D spiral | 0.99 | 0.99 | 11.6 | 9.0 |
| 3D radial | 0.99 | 0.99 | 15.3 | 10.5 |
| 3D spiral | 0.98 | 0.98 | 7.0 | 4.9 |
| | | T1 3T | | |
| | CCC SVD | CCC KW-SVD | NRMSE SVD [%] | NRMSE KW-SVD [%] |
| 2D radial | 0.99 | 0.98 | 8.3 | 10.1 |
| 2D spiral | 0.99 | 0.98 | 6.2 | 10.4 |
| 3D radial | 0.99 | 0.99 | 9.5 | 8.7 |
| 3D spiral | 0.99 | 0.99 | 7.7 | 6.4 |
| **T2 3T** | | | | |
| | CCC SVD | CCC KW-SVD | NRMSE SVD [%] | NRMSE KW-SVD [%] |
| 2D radial | 0.99 | 0.99 | 13.5 | 10.1 |
| 2D spiral | 0.92 | 0.94 | 13.5 | 12.7 |
| 3D radial | 0.99 | 0.99 | 13.7 | 11.0 |
| 3D spiral | 0.97 | 0.98 | 9.6 | 7.6 |

*Table 3: **In vivo results using the NN multipath architecture.** Results for representative ROIs within white matter (WM) and gray matter (GM).*

| | | 1.5 T | | | |
|---|---|---|---|---|---|
| | | T1 WM [ms] | T2 WM [ms] | T1 GM [ms] | T2 GM [ms] |
| **Previous reports**[49–51] | | 788-898 | 63-80 | 1269-1393 | 78-117 |
| **2D radial** | SVD | 600 ± 221 | 53 ± 20 | 1189 ± 239 | 92 ± 32 |
| | KW-SVD | 731 ± 67 | 64 ± 8 | 1046 ± 90 | 89 ± 14 |
| **2D spiral** | SVD | 677 ± 96 | 56 ± 12 | 1041 ± 243 | 92 ± 30 |
| | KW-SVD | 774 ± 49 | 64 ± 7 | 1004 ± 104 | 87 ± 15 |
| **3D radial** | SVD | 699 ± 225 | 64 ± 24 | 1337 ± 422 | 132 ± 109 |
| | KW-SVD | 688 ± 117 | 63 ± 13 | 1381 ± 304 | 112 ± 17 |
| **3D spiral** | SVD | 647 ± 60 | 65 ± 23 | 1189 ± 119 | 101 ± 23 |
| | KW-SVD | 630 ± 54 | 65 ± 7 | 1192 ± 90 | 100 ± 14 |

*Table 4: Reconstruction and inference times.* Measurements were performed on an Intel Xeon processor E5-2600 v4 (48 CPU cores) equipped with a NVIDIA Tesla K80 GPU. the inference times are compared between exhaustive grid search (dictionary matching), fast group matching, and neural network. 2D results are reported for a single slice of the spiral trajectory.

| Reconstruction time [s] | | | |
|---|---|---|---|
| 2D | CPU | zero-filling | 2.2 |
| | | view-sharing | 5.8 |
| | GPU | zero-filling | 0.6 |
| | | view-sharing | 4 |
| 3D | CPU | zero-filling | 250 |
| | | view-sharing | 423 |
| | GPU | zero-filling | 160 |
| | | view-sharing | 357 |

| Inference time [s] | | | |
|---|---|---|---|
| 2D | CPU | Dictionary | 9.5 |
| | | Group matching | 0.6 |
| | | Neural network | 0.1 |
| | GPU | Dictionary | 6.6 |
| | | Group matching | 0.5 |
| | | Neural network | <0.1 |
| 3D | CPU | Dictionary | 618 |
| | | Group matching | 90 |
| | | Neural network | 10 |
| | GPU | Dictionary | 115 |
| | | Group matching | 68 |
| | | Neural network | 10 |

**Table 5: Previous reports of 3D MRF acquisitions as compared to our 3D QTI proposal.**

|  | Ma et al.[10] | Ma et al.[12] | Liao et al.[7] | Chen et al.[36] | Cao et al.[11] | Proposed |
|---|---|---|---|---|---|---|
| **Trajectory** | Stack-of-Spirals | Music Generated | Stack-of-Spirals | Stack-of-Spirals | Spiral Projections | Spiral Projections |
| **Resolution** | 1.2x1.2x3 mm³ | 2.3 isotropic | 1mm isotropic | 1mm isotropic | 1/0.8mm isotropic | 0.96 isotropic |
| **Spatial coverage** | 300x300x144 mm³ | 300x300x300 mm³ | 260x260x192 mm³ | 260x260x180 mm³ | 240x240x240 mm³ | 192x192x192 mm³ |
| **Acquisition time** | 4.6 min | 37 min | 7.5 min | 6.5 - 7.1 min | 5/6 min | <5 min |
| **Reconstruction time** | 48 min | N.A. | 20 h | N.A. | 1.5-4h | < 7 min |